\documentclass{blois}

\def\be{\begin{equation}}
\def\ee{\end{equation}}
\def\bea{\begin{eqnarray}}
\def\eea{\end{eqnarray}}

\def\bc{\begin{center}}
\def\ec{\end{center}}

\newcommand{\Xt}{\tilde{X}}
\newcommand{\yl}{{Y^{\ast}_L}}
\newcommand{\yr}{{Y^\ast_{R}}}
\newcommand{\nc}{\newcommand}
\nc{\vp}{\varphi}
\nc{\tvp}{\widetilde{\varphi}}
\nc{\vpj }{\mbox{${\vp^\dag i\,\raisebox{2mm}{\boldmath ${}^\leftrightarrow$}\hspace{-4mm} D_\mu\,\vp}$}}
\nc{\vpjt}{\mbox{${\vp^\dag i\,\raisebox{2mm}{\boldmath ${}^\leftrightarrow$}\hspace{-4mm} D_\mu^{\,I}\,\vp}$}}

\newcommand{\Photo}{}

\begin{document}
\vspace*{4cm}
\title{THEORETICAL ASPECTS OF FLAVOUR AND CP VIOLATION IN THE LEPTON SECTOR}

\author{ FERRUCCIO FERUGLIO }

\address{Dipartimento di Fisica e Astronomia `G.~Galilei', Universit\`a di Padova
\\
INFN, Sezione di Padova, Via Marzolo~8, I-35131 Padua, Italy}

\maketitle\abstracts{
We review flavour and CP violations in the lepton sector as probes of new physics beyond the 
standard model and its minimal extensions accommodating massive neutrinos. After recalling the main experimental bounds
and the future perspectives, we summarize the limits on a set of dimension-six gauge-invariant operators, pointing to a scale of new physics
much larger than the electroweak scale. If we insist on having new physics at the TeV scale, as demanded by most of the standard model extensions 
addressing the gauge hierarchy problem, we should rely on special mechanisms to deplete flavour and CP violations in the charged lepton sector.
We comment on the capability of partial compositeness in reconciling the absence of signals with a scale of new physics accessible to 
the LHC.}

\section{Motivations and experimental searches}
Lepton flavour violation in the charged lepton sector (CLFV) is expected at some level. The individual lepton numbers $L_i$ $(i=e,\mu,\tau)$ are violated 
in neutrino oscillations. We have evidence for conversion of $\nu_e$ into $(\nu_\mu,\nu_\tau)$ from solar neutrino experiments, $\nu_\mu$ into $\nu_e$ from long baseline experiments
and $\nu_\mu$ into $\nu_\tau$ from atmospheric neutrino oscillations and the OPERA experiment \cite{Altarelli:2014dca}. Massive neutrinos and a non-trivial lepton mixing matrix $U_{PMNS}$
imply $L_i$ $(i=e,\mu,\tau)$  non-conservation, which should show up in processes with charged leptons \footnote{The total lepton number $L=L_e+L_\mu+L_\tau$ can be (classically) conserved or not, depending of the type of neutrino masses.}.
The violation of the individual lepton numbers in the context of three generations entails CP violation in the lepton sector (LCPV). This is welcome since the amount of CP violation from the quark sector
is insufficient to generate the baryon asymmetry of the universe. For example in the standard model (SM) minimally extended by the inclusion of three right-handed neutrinos ($\nu$SM) there are six independent CP-violating phases,
leading to the attractive possibility of leptogenesis. 

At the level of the present and forthcoming experimental sensitivities, both CLFV and LCPV probe new physics (NP) beyond the $\nu$SM. For instance the branching ratio of the radiative muon decay
$\mu\to e \gamma$, evaluated in the $\nu$SM assuming Dirac neutrino masses, is given by
\be
BR(\mu\to e \gamma)\approx\frac{3\alpha}{32\pi}\left\vert U^*_{\mu i} U_{ei}\frac{m_i^2}{m_W^2} \right\vert^2\approx 10^{-53}~~~,
\ee
completely out of reach by conceivable experiments by many orders of magnitude. This process is unobservable also in $\nu$SM within a large scale type-I see-saw.
The smallness of $BR(\mu\to e \gamma)$ is explained by the peculiar GIM suppression in the above expression: in the quark case we have small mixing angles and large masses, while here we have relatively large angles but tiny neutrino masses.
In the SM with massless neutrinos the electric dipole moment (EDM) of the electron gets the first non-vanishing contribution at the 4th loop order:
\be
\frac{d_e}{e}\approx \frac{G_F m_e}{\pi^2}\left(\frac{\alpha}{2\pi}\right)^3 J\approx 6\times 10^{-37}~cm~~~~~~~~~~~~~~\left[J=(2.96^{+0.20}_{-0.16})\times 10^{-5}\right]~~~.
\ee
In the $\nu$SM with Dirac (Majorana) neutrinos $d_e$ is different from zero at 3(2) loops, but is negligibly small \cite{Archambault:2004td}.
Any evidence for CLFV and/or LCPV would imply NP beyond the $\nu$SM. 

The absence of signals of CLFV and LCPV is one of the aspects of the more general flavour problem, which
follows from the assumption that the gauge hierarchy problem is solved by NP close to the TeV scale.
Indeed, given the current sensitivity to CLFV and LCPV processes, a NP scale $\Lambda_{\rm NP}$ much larger than the electroweak scale is naively expected.
Thus, if there is NP at the TeV scale as demanded by most of the SM extensions addressing the gauge hierarchy problem, we should devise special mechanisms
to deplete CLFV and LCPV much more efficiently than the suppression factor $E^2/\Lambda_{NP}^2$ carried by the relevant amplitudes.
Another aspect we would like to clarify is the link between CLFV, LCPV and neutrino properties.
 
\begin{table}[t]
\caption[]{Upper bounds on the branching ratios of some LFV tau decays.}
\label{tab0}
\vspace{0.4cm}
\begin{center}
\begin{tabular}{|l|l|l|l|}
\hline
&Present upper bound&&Present upper bound\\
\hline
BR$(\tau\to e\gamma)$& $3.3\times 10^{-8}$&BR$(\tau\to 3e)$& $2.7\times 10^{-8}$\\
\hline
BR$(\tau\to \mu\gamma)$& $4.4\times 10^{-8}$&BR$(\tau\to 3\mu)$& $2.1\times 10^{-8}$\\
\hline
\end{tabular}
\end{center}
\end{table}

 \begin{table}[h]
\caption[]{Present upper bounds and future expected sensitivity on the branching ratios of LFV muon decays.}
\label{tab1}
\vspace{0.4cm}
\begin{center}
\begin{tabular}{|l|l|l|}
\hline
&Present upper bound & Future sensitivity \\
\hline
BR$(\mu^+\to e^+\gamma)$& $5.7\times 10^{-13}$~~~[MEG]&  $6\times 10^{-14}$~~~[MEG 2018]\\
\hline
BR$(\mu^+\to e^+e^+e^-)$& $1.0\times 10^{-12}$~~~[SINDRUM]&  $\approx10^{-16}$~~~[Mu3e$>$2019]\\
\hline
CR$(\mu^- Ti\to e^- Ti)$& $4.3\times 10^{-12}$~~~[SINDRUM II]&  \\
\hline
CR$(\mu^- Au\to e^- Au)$& $7.0\times 10^{-13}$~~~[SINDRUM II]&  \\
\hline
CR$(\mu^- Al\to e^- Al)$& & $(2\div 6)\times 10^{-17}$~~~[Mu2e$>$2018] \\
\hline
CR$(\mu^- Al\to e^- Al)$& & $\approx3\times 10^{-17}$~~~[COMET$>$2018] \\
\hline
\end{tabular}
\end{center}
\end{table}

The experimental searches of CLFV cover tau and muon decays and $\mu$ to $e$ conversion in nuclei. In tau decays we have a rich pattern of kinematically allowed channels, ranging from $\tau\to \mu\gamma$, $\tau\to e \gamma$
to tau decays into three charged leptons and LFV semileptonic tau decays. The current limits on the corresponding BRs are at the level of few $10^{-8}$, at 90\% C.L., from the searches carried out at the 
BABAR and Belle B-factories, see table \ref{tab0}. The future expected sensitivity should go down to the $10^{-9}$ level for most of these channels, from the searches planned at future super B-factories and LHCb. 
The muon is the major player in this field. The most stringent limits on NP come from the present bounds on LFV muon decays. Great improvements are expected within this decade,
at the level of $4\div 5$ orders of magnitude for the sensitivities of $\mu \to 3 e$ and $\mu \to e$ conversion in nuclei, see table \ref{tab1}: we can speak of a golden age for CLFV searches.
Concerning the anomalous magnetic moments and the EDM of the charged leptons, the present data are summarized in table \ref{tab2}. We recall the long-standing evidence ($\approx 3.2\sigma$) of a deviation
with respect to the SM prediction in $a_\mu=(g-2)_\mu/2$, which will be checked by the Muon g-2 experiment at Fermilab, aiming to bring the present 0.5 ppm accuracy down to the 0.2 ppm level.
\begin{table}[h!]
\caption[]{Present bounds/results on the EDM and anomalous magnetic moments of the charged leptons. }
\label{tab2}
\vspace{0.4cm}
\begin{center}
\begin{tabular}{|l|l|l|}
\hline
$l$& $d_l$~($e~cm$)& $\Delta a_l=a_l^{EXP}-a_l^{SM}$\\
\hline
$e$&$<8.7\times 10^{-29}$&$(-10.5\pm 8.1)\times 10^{-13}$\\
\hline
$\mu$&$<1.8\times 10^{-19}$&$(29\pm 9)\times 10^{-10}$\\
\hline
$\tau$&$<10^{-16}$&$-0.007<\Delta a_\tau<0.005$\\
\hline
\end{tabular}
\end{center}
\end{table}
\section{Effective Lagrangian and model-independent bounds}
To describe generic NP contributions to CLFV and LCPV it is convenient to adopt an effective field theory description where the SM Lagrangian is extended
by a set of gauge invariant operators depending on the SM fields \cite{Buchmuller:1985jz}:
\be
\mathcal{L} = \mathcal{L}_{\rm SM} + 
\frac{1}{\Lambda_{NP}^2} \sum_i C_i\, Q_i +...
\label{Lag6}
\ee
\begin{table}[t]
\caption[]{Dimension-six operators relevant to the present discussion. $W_{\mu\nu}^I$ and $B_{\mu\nu}$ denote the field strengths for the gauge vector bosons of SU(2) and U(1), respectively; 
$\theta_W$ is the weak mixing angle and $Q_{eW3}$ is the contribution to $Q_{eW}$ obtained setting to zero $W_{\mu\nu}^{1,2}$.}
\label{tab3}
\vspace{0.4cm}
\begin{center}
\begin{tabular}{|c|c|c|}
\hline
&$(Q_{eW})_{ij}$&$(\bar \ell_{Li} \sigma^{\mu\nu} \tilde e_{Rj}) \tau^I \vp W_{\mu\nu}^I$\\
&$(Q_{eB})_{ij}$&$(\bar \ell_{Li}  \sigma^{\mu\nu} \tilde e_{Rj}) \vp B_{\mu\nu}$\\
{\tt Dipole}&$(Q_{e\gamma})_{ij}$&$\cos\theta_W~(Q_{eB})_{ij}-\sin\theta_W~(Q_{eW3})_{ij}$\\
&$(Q_{eZ})_{ij}$&$\sin\theta_W~(Q_{eB})_{ij}+\cos\theta_W~(Q_{eW3})_{ij}$\\
\hline
\hline
&$(Q_{\vp l}^{(1)})_{ij}$&$(\vpj)( \bar \ell_{Li}\gamma^\mu \ell_{Lj})$\\
{\tt Vector}&$(Q_{\vp l}^{(3)})_{ij}$&$(\vpjt)(\bar \ell_{Li} \tau^I \gamma^\mu \ell_{Lj})$\\
&$(Q_{\vp e})_{ij}$&$(\vpj)(\overline{\tilde e}_{Ri} \gamma^\mu \tilde e_{Rj})$\\
\hline
\hline
{\tt Scalar}&$(Q_{e\vp})_{ij}$&$(\vp^\dag \vp)(\bar \ell_{Li} \tilde e_{Rj} \vp)$\\
\hline
\hline
&$(Q_{ll})_{ijmn}$&$(\bar \ell_{Li} \gamma_\mu \ell_{Lj})(\bar \ell_{Lm} \gamma^\mu \ell_{Ln})$\\
{\tt Contact}&$(Q_{ee})_{ijmn}$&$(\overline{\tilde e}_{Ri} \gamma_\mu \tilde e_{Rj})(\overline{\tilde e}_{Rm} \gamma^\mu \tilde e_{Rn})$\\
&$(Q_{le})_{ijmn}$&$(\bar \ell_{Li} \gamma_\mu \ell_{Lj})(\overline {\tilde e}_{Rm} \gamma^\mu \tilde e_{Rn})$\\
\hline
\end{tabular}
\end{center}
\end{table}
where we have focused our attention on the dimension-six operators, which are the lowest-dimensional ones contributing to the processes of interest.
Dots stand for higher-dimensional operators. A complete set of dimension-six operators \cite{Grzadkowski:2010es}  depending on lepton fields and on the scalar electroweak 
doublet $\vp$ is given in table \ref{tab3}.
The electromagnetic dipole operators $(Q_{e\gamma})_{ij}$
are the only operators that give a tree-level contribution to the radiative decays of charged leptons, when $i\ne j$. The diagonal elements, $i=j$, contribute
to the anomalous magnetic moments and to the EDM of the charged leptons. After the breaking of the electroweak symmetry the vector operators $(Q_{\vp l}^{(1,3)})_{ij},(Q_{\vp e})_{ij}$ modify the couplings of the $Z$ boson to leptons, violating both universality and lepton flavour. The scalar operators $(Q_{e\vp})_{ij}$
contribute, with a different weight, to masses and Higgs couplings of the charged leptons. The four-lepton operators $(Q_{ll})_{ijmn},(Q_{ee})_{ijmn},(Q_{le})_{ijmn}$
can contribute to muon and tau decays into three charged leptons. Other ten independent dimension-six operators of the type $\ell\ell qq$ describe $\mu$ to $e$ conversion in nuclei \cite{Grzadkowski:2010es}.
Each operator carries flavour indices and hermiticity of the Lagrangian is guaranteed either by appropriate symmetry properties of the
coefficients under transposition of the indices or by addition of the hermitian conjugate operator. 
\begin{table}[h!]
\caption{Bounds on off-diagonal Wilson coefficients 
$C_{A}^{ij}/\Lambda_{NP}^2$ ($A={\tt Dipole, Scalar, Vector}$). 
In the second column we list the upper bound on $|C_{A}^{ij}|$
assuming $\Lambda_{NP}=1$ TeV, while in the third column we fix $|C_{A}^{ij}|=1$ and we list the corresponding lower bound on $\Lambda_{NP}$, in TeV. Bounds on the coefficients $C_{A}^{ji}$ are equal to the bounds on the coefficients $C_{A}^{ij}$.}
\label{t4}
\vspace{0.4cm}
\begin{center}
\begin{tabular}{| l | c | c | l |}
\hline
 & $|c|$ ($\Lambda_{NP}=1$ TeV) & $\Lambda_{NP}$ (TeV) ( $|c| =1$) & \\
 \hline
 \hline
 $C_{e\gamma}^{\mu e}$& $2.5\times 10^{-10}$ & $6.3\times 10^{4}$ & $\mu\to e \gamma$\\
 \hline
$C_{e\gamma}^{\tau e}$& $2.4\times 10^{-6}$ & $6.5\times 10^{2}$ & $\tau\to e \gamma$\\
 \hline
 $C_{e\gamma}^{\tau\mu}$& $2.7\times 10^{-6}$ & $6.1\times 10^{2}$ & $\tau\to \mu \gamma$\\
 \hline
$C_{eZ}^{\mu e}$& $1.4\times 10^{-7}$ & $2.7\times 10^{3}$ & $\mu\to e \gamma$ {[\tt 1-loop]}\\
 \hline
$C_{eZ}^{\tau e,\tau\mu}$& $\approx\times 10^{-3}$ & $\approx 30$ & $\tau\to e \gamma$, $\tau\to \mu \gamma$ {[\tt 1-loop]}\\
\hline
\hline
$C_{e\varphi}^{\mu e}$& $8.4\times 10^{-5}$ & $109$ & $\mu\to e \gamma$ {[\tt 2-loop]}\\
 \hline
$C_{e\varphi}^{\tau e}$& $0.33$ & $1.7$ & $\tau\to e \gamma$ {[\tt 2-loop]}\\
 \hline
 $C_{e\varphi}^{\tau\mu}$& $0.37$ & $1.6$ & $\tau\to \mu \gamma$ {[\tt 2-loop]}\\
 \hline
 \hline
$(C_{\vp l}^{(1,3)})_{\mu e}$, $C_{\vp e}^{\mu e}$& $4\times 10^{-5}$ & $160$ & $\mu\to 3 e$\\
  \hline
$(C_{\vp l}^{(1,3)})_{\tau e}$, $C_{\vp e}^{\tau e}$& $1.5\times 10^{-2}$ & $\approx 8$ & $\tau\to 3 e$\\
  \hline
$(C_{\vp l}^{(1,3)})_{\tau\mu}$, $C_{\vp e}^{\tau\mu}$& $\approx 10^{-2}$ & $\approx 9$ & $\tau \to 3\mu$\\
  \hline
 \end{tabular}
\end{center}
\end{table}

In tables \ref{t4} and \ref{t7} we list the current bounds on the Wilson coefficients $C_i$ of the Lagrangian of eq.~(\ref{Lag6}), extracted from
model-independent studies of LFV processes \cite{Brignole:2004ah,Crivellin:2013hpa,Pruna:2014asa}.
Each bound has been derived by assuming a single non-vanishing Wilson coefficient at the time. 
The main bounds on the coefficients of the dipole operators $Q_{e\gamma}$ and $Q_{eZ}$ come from the present limits on the branching ratios of the radiative lepton decays.
The amplitudes for these processes get a tree-level contribution from the off-diagonal elements of the electromagnetic dipole operator $Q_{e\gamma}^{ij}$, while $Q_{eZ}^{ij}$ 
contributes at one loop. 
From the limits on the lepton EDMs we have  \cite{Crivellin:2013hpa}
\be
Im(C_{e\gamma}^{ee})\left(\frac{1~{\rm TeV}}{\Lambda_{NP}}\right)^2<3.9 \times 10^{-12}~~~,~~~~~~~
Im(C_{e\gamma}^{\mu\mu})\left(\frac{1~{\rm TeV}}{\Lambda_{NP}}\right)^2<8.1 \times 10^{-3}~~~.
\label{edm}
\ee
Given the current deviation $\Delta a_\mu=a_\mu^{EXP}-a_\mu^{SM}$ in the muon anomalous magnetic moment \cite{Passera:2004bj,Jegerlehner:2009ry} $a_\mu = (g-2)_\mu /2$  reported in table \ref{tab2}
we would need
\be
Re(C_{e\gamma}^{\mu\mu})\left(\frac{1~{\rm TeV}}{\Lambda_{NP}}\right)^2=1.2\times 10^{-5}~~~,
\label{gm2}
\ee 
to reproduce the central value of the data.

Also the scalar operator $Q_{e\varphi}$ is mostly bounded by  the limits on radiative lepton decays \cite{Goudelis:2011un,Harnik:2012pb,Blankenburg:2012ex,Crivellin:2013hpa,Pruna:2014asa}. 
These bounds are dominated by two-loop contributions of the corresponding operator to the radiative lepton decay,
through Barr-Zee type diagrams, assuming that the top Yukawa coupling is as in the SM. One-loop contributions to charged lepton radiative decays and tree-level contributions to $l\to 3l'$ decays lead to less severe bounds than the ones given in table \ref{t4}.

\begin{table}[h!]
\caption{Bounds on  coefficients $C_{ll,ee,le}^{ijkl}$.
In the second column we list the upper bound on the Wilson coefficients
assuming $\Lambda_{NP}=1$ TeV, while in the third column we set to unity the coefficients and we list the corresponding lower bound on $\Lambda_{NP}$, in TeV.}
\label{t7}
\vspace{0.4cm}
\begin{center}
\begin{tabular}{| c | c | c | c |}
\hline
 & $|c|$ ($\Lambda_{NP}=1$ TeV) & $\Lambda_{NP}$ (TeV) ( $|c| =1$) & \\
  \hline
$C_{ll,ee}^{\mu eee}$& $2.3\times 10^{-5}$ & $207$ & $\mu\to 3 e$\\
  \hline
$C_{ll,ee}^{e\tau ee}$& $9.2\times 10^{-3}$ & $10.4$ & $\tau\to 3 e$\\
  \hline
$C_{ll,ee}^{\mu\tau \mu\mu}$& $7.8\times 10^{-3}$ & $11.3$ & $\tau \to 3\mu$\\
  \hline
$C_{le}^{\mu eee, e e\mu e}$& $3.3\times 10^{-5}$ & $174$ & $\mu\to 3 e$\\
  \hline
  $C_{le}^{\mu \mu\mu e, e \mu \mu\mu}$& $2.1\times 10^{-4}$ & $69$ & $\mu\to e\gamma$ {[\tt 1-loop]} \\
  \hline
  $C_{le}^{\mu \tau\tau e, e \tau\tau\mu}$& $1.2\times 10^{-5}$ & $289$ & $\mu\to e\gamma$ {[\tt 1-loop]}\\
  \hline
$C_{le}^{e\tau ee,e e e\tau}$& $1.3\times 10^{-2}$ & $8.8$ & $\tau\to 3 e$\\
  \hline
$C_{le}^{\mu\tau\mu\mu,\mu\mu\mu\tau}$& $1.1\times 10^{-2}$ & $9.5$ & $\tau \to 3\mu$\\
  \hline
\end{tabular}
\end{center}
\end{table}
Coming to the vector operators $(Q_{\vp l}^{(1,3)})_{ij}$, $(Q_{\vp e})_{ij}$, they lead to lepton flavour violating $Z$ decays, but the corresponding limits on the Wilson coefficients,
assuming $\Lambda_{NP}=1$ TeV, are of order 10\% \cite{Crivellin:2013hpa}. Through one-loop diagrams they also contribute to radiative decays of the charged leptons \cite{Crivellin:2013hpa,Pruna:2014asa}.
It turns out that the most restrictive bounds come from the decays $l\to 3l'$, whose branching ratios satisfy the limits given in tables \ref{tab0} and \ref{tab1}.
We collect the corresponding bounds in table \ref{t4}.
Also the contact operators can contribute to both the decays $l\to 3l'$ and, through one-loop diagrams, to the radiative decays of the charged leptons. 
The most significant bounds are given in table \ref{t7}. Finally, present limits on $\mu$ to $e$ conversion in nuclei constrain the Wilson coefficients of operators of the type $\ell\ell qq$ \cite{Feruglio:2015gka}.
\section{A closer look to the dipole operator}
In the electromagnetic dipole operator
\be
\frac{C_{e\gamma}^{ij}}{\Lambda_{NP}^2}(\bar \ell_{Li}  \sigma^{\mu\nu} \tilde e_{Rj}) \vp F_{\mu\nu}~~~,
\ee
it is convenient to redefine the Wilson coefficient as
\be
{C_{e\gamma}^{ji}}^*=e A_{ij}\frac{m_{[ij]}}{\sqrt{2}v}~~~~~~~~~~~~~~~[ij]=max(ij)~~~,
\label{redef}
\ee
to account for both the electromagnetic coupling constant $e$ and for the violation of the chiral symmetry, since we expect that $C_{e\gamma}^{ij}$ should vanish in the chiral limit.
Considering $i=j$ we have
\be
\Delta a_i=2\frac{m_i^2}{\Lambda_{NP}^2}Re(A_{ii})~~~,~~~~~~~\frac{d_i}{e}=\frac{m_i}{\Lambda_{NP}^2} Im(A_{ii})~~~,
\label{ad}
\ee
implying the model-independent relation
\be
\frac{d_i}{e}=\frac{\Delta a_i}{2 m_i}\tan\varphi_i~~~,~~~~~~~~~\varphi_i=Arg(A_{ii})~~~.
\label{modind}
\ee
Taking $A_{\mu\mu}$ of order 1, we need $\Lambda_{NP}\approx 3$ TeV, to account for the central value of the discrepancy in $a_\mu$, see table \ref{tab2}. From eq. (\ref{modind}) we get
\be
\frac{d_\mu}{e}\approx 3\times 10^{-22}\left(\frac{\Delta a _\mu}{30\times 10^{-10}}\right)~\tan\varphi_\mu~~~cm     ~~~,              
\ee
much smaller than the current bound if $\varphi_\mu$ is of order one. The assumption $A_{ii}=A$ defines the so-called limit of Na\"ive Scaling (NS), where the following relations hold
\be
\frac{\Delta a_i}{\Delta a_j}=\frac{m^2_i}{m^2_j}~~~,~~~~~~~\frac{d_i}{d_j}=\frac{m_i}{m_j}~~~.
\ee 
NS is a useful benchmark, but it can be violated in SM extensions with new flavoured particles having non-universal masses and/or non-universal interactions with leptons \cite{Giudice:2012ms}.
According to NS we have
\be
\Delta a_e=\frac{m^2_e}{m^2_\mu}~\Delta a_\mu\approx 7\times 10^{-14}\left(\frac{\Delta a _\mu}{30\times 10^{-10}}\right)~~~.
\label{aeNS}
\ee
The present experimental sensitivity to $\Delta a _e$
\be
\Delta a_e=a_e^{EXP}-a_e^{SM}=(-10.5\pm 8.1)\times 10^{-13}~~~,
\label{ae}
\ee
is only one order of magnitude far from the prediction of eq. (\ref{aeNS}), as pointed out in ref. \cite{Giudice:2012ms}. In eq. (\ref{ae}) the error is dominated by the experimental error on $a_e^{EXP}$ and by the 
error on the fine structure constant $\alpha(^{78}Rb)$ as extracted from the Rydberg constant through atom interferometry. A realistic reduction of these two errors, probably
achievable in the near future, would allow to verify the relation between $\Delta a_\mu$ and $\Delta a_e$. 

If NS holds we also have
\be
\frac{d_e}{e}=\frac{1}{2m_e}\frac{m_e^2}{m_\mu^2}\Delta a_\mu \tan\varphi\approx 1.4\times 10^{-24}\left(\frac{\Delta a _\mu}{30\times 10^{-10}}\right)\tan\varphi~cm~~~,
\ee
showing that the $(g-2)_\mu$ anomaly and the NS require tiny flavour-blind phases, $|\varphi|<6\times 10^{-5}$.
\section{Why we have not seen CLFV and LCPV?}
An efficient mechanism suppressing flavour changing neutral currents (FCNC) and CP violation
can be introduced by observing that in the electroweak theory the symmetry of the flavour sector is broken only by the Yukawa interactions. 
Minimal Flavour Violation (MFV) \cite{D'Ambrosio:2002ex} is defined by the assumption that, even including NP contributions, Yukawa couplings are the only source of such symmetry breaking.
In MFV flavour effects from NP are controlled and damped by the smallness of fermion masses and mixing angles. In this framework data from the quark sector 
allow $\Lambda_{NP}$ to be considerably smaller, close to the TeV scale. In the lepton sector MFV is not unambiguously defined \cite{Cirigliano:2005ck}, 
since we have several different ways to describe neutrino masses. For instance  we can add to the SM Yukawa Lagrangian the dimension-five gauge-invariant Weinberg operator:
\be
{\cal L}_Y=-\bar \ell_{Li}  \vp~\hat y_{\ell ii}~ \tilde e_{Ri}- \frac{1}{2\Lambda_L} \bar \ell_{Li}  \vp ~w_{ij}~ \bar \ell_{Lj}  \vp+h.c.
\ee
Here we work in the basis where $\hat y_\ell$, the matrix of charged lepton Yukawa couplings, is diagonal. The matrix $w_{ij}$ describes neutrino masses and lepton mixing angles and the 
scale $\Lambda_L$ is associated to the breaking of the total lepton number. Assuming MFV, we can estimate the Wilson coefficients of the dimension-six operators.  
For instance, for the electromagnetic dipole operators $(Q_{e\gamma})_{ij}$ we have:
\be
C^{ij}_{e\gamma}=\left[(c_1~\mbox{l\hspace{-0.55em}1}+c_2~ \hat y_\ell \hat y_\ell^\dagger + c_3~ w w^\dagger+....)~\hat y_\ell\right]_{ij}~~~~,
\ee
where $c_i$ are real coefficients of order one, dots stand for insertions of higher order in $(\hat y_\ell,w)$ and the lowest order LFV contribution comes from the term proportional to $c_3$. As in the SM, in the limit of vanishing neutrino masses $w=0$
there is no LFV. This is more evident when we express the LFV part of $C_{e\gamma}^{ij}$ in terms of neutrino masses and lepton mixing angles:
\be
w w^\dagger \hat y_\ell=\frac{4\sqrt{2}}{v^3}\frac{\Lambda_L^2}{v^2}~U_{PMNS}~\hat m_\nu^2~ U_{PMNS}^\dagger~ \hat m_e~~~.
\label{c3}
\ee
The observability of radiative charged lepton decays depends on the ratio $\Lambda_L/\Lambda_{NP}$. All experimental bounds are satisfied if $\Lambda_L/\Lambda_{NP}<10^9$.
Qualitatively similar conclusions hold if neutrino masses are described by a type-I see-saw mechanism. 
In the ratio $BR(\mu\to e \gamma)/BR(\tau\to\mu\gamma)$ the unknown scales $\Lambda_{NP,L}$ drop
and we get relatively accurate predictions. In particular, from the present limit on $BR(\mu\to e \gamma)$ we find $BR(\tau\to\mu\gamma)<(1.0\div1.6)\times 10^{-11}$, the source of uncertainty being
the mixing matrix $U_{PMNS}$ and in particular the Dirac-like CP-violating phase.
MFV provides a useful benchmark for the discussion of the flavour sector, but it does not emerge
from the known mechanisms aiming to explain the observed fermion spectrum \cite{Feruglio:2015jfa}, or from the known models providing a solution to the gauge hierarchy problem.

It is interesting to see what is the degree of suppression of CLFV and LCPV achievable in SM extensions addressing the gauge hierarchy problem. Beyond low-energy supersymmetry,
where the predictions are rather model dependent and strictly related to the mechanism which breaks supersymmetry, a class of attractive candidates is provided by composite Higgs models \cite{Bellazzini:2014yua,Panico:2015jxa},
where fermion masses and mixing angles can be described by partial compositeness \cite{Kaplan:1991dc}. Light fermions get hierarchical masses from the mixing between an elementary sector and a composite one. 
As a toy realization of this idea, consider a model where
the composite sector contains, for each SM fermion, a pair of heavy fermions allowing a Dirac mass term of the order of the compositeness scale and a 
mixing term with the SM fields \cite{Contino:2006nn}
\bea
{\cal L}_Y&=& - \sum\limits_{i,j=1}^{3} \left( \bar\ell_{Li} \Delta_{ij} L_{Rj} - 
      \bar{\tilde{e}}_{Ri} \tilde{\Delta}_{ij} \tilde{E}_{Lj} \right) + h.c.\\
&& -\sum\limits_{i=1}^{3} \left( \bar{L}_{i} m_i L_{i}+ 
      \bar{\tilde{E}}_{i} \tilde{m}_i \tilde{E}_{i} \right) \\
&& - \sum\limits_{i,j=1}^{3} \left(\bar{L}_{Ri}\varphi \yl_{ij} \tilde{E}_{Lj} +
      \bar{L}_{Li}\varphi\yr_{ij}\tilde{E}_{Rj} \right) + h.c.
\eea
The first line represents the mixing between the elementary sector and the composite one, the second line displays Dirac mass terms for the fermions of the composite
sector and the third line shows the Yukawa interactions that are restricted to the composite sector alone and are assumed to lie in the range
$4\pi\ge |\yr|,|\yl|\ge 1$. With the above Lagrangian neutrinos are massless and we can analyze CLFV and LCPV in this limit.
By integrating out the composite sector under the assumptions $m_i=m$, $\tilde m_i=\tilde m$ and $\tilde m,m\gg v$, we get the SM-like Yukawa interaction
\be
{\cal L}_Y^{eff}=-\bar \ell_{L}  \vp~\hat y^{SM}_{\ell}~ \tilde e_{R}+...~~~~~~~~~~~~~~~~~~~~~~y^{SM}_\ell = X \yr \tilde X^\dagger~~~,
\ee
where $X\equiv\Delta m^{-1}$, $\tilde X\equiv\tilde\Delta \tilde m^{-1\dagger}$ and dots stand for contributions of higher order in $v/m$. The appealing feature of this pattern is that 
hierarchical fermion masses and mixing angles can be explained by the mixing matrices $X$ and $\tilde X$, even in the presence of 
anarchical matrices $\yr$ and $\yl$. This is of particular interest, since neutrino masses and mixing angles as extracted from neutrino 
oscillation experiments \cite{Altarelli:2014dca} seem to support the idea of an underlying anarchical dynamics.
At the one-loop level, summing over the $h,Z$ and $W$ amplitudes, we get the main contribution to the electromagnetic dipole operator $Q_{e\gamma}$:
\be
\frac{\left(C_{e\gamma}\right)_{h+Z+W}}{\Lambda_{NP}^2} = \frac{e}{64\pi^2} \frac{1}{m \tilde m} X \yr \yl^\dagger \yr \Xt^\dagger\;.
\label{dipole}  
\ee
If $\yr$ and $\yl$ are anarchical matrices we see that, in general, $y^{SM}_\ell$ and $C_{e\gamma}$ are not diagonal in the same basis. At variance with MFV, we have
CLFV even in the limit of vanishing neutrino masses, without any relation to the scale of breaking of the total lepton number $L$. Moreover the coefficients $(C_{e\gamma})_{ij}$
are in general complex in the basis where $y^{SM}_\ell$ is real, positive, diagonal and CP violation is induced.
From the present bounds on $BR(\mu\to e \gamma)$ and $d_e$ we find that, with reasonable assumptions on the mixing matrices $X$ and $\tilde X$,
we need $m/\langle Y\rangle$ and $\tilde m/\langle Y\rangle$ well above 10 TeV, $\langle Y\rangle$ denoting an average absolute value of the Yukawa matrices.
Even though in the presence of anarchical Yukawa interactions the degree of suppression provided by partial compositeness does not allow for a compositeness scale
$m,\tilde m$ of the order of 1 TeV, we should stress the improvement by several orders of magnitude with respect to the bounds of the general effective operator analysis presented in the previous section.

Assuming that the left mixing $X$ is proportional to the identity, as suggested by the large mixing angles in the lepton sector, is not
sufficient to align $y^{SM}_\ell$ and $C_{e\gamma}$. If we postulate that $\yl=0$, charged lepton masses do not vanish while the leading order contribution (\ref{dipole})
does. In this case we should pay attention to NLO contributions to $Q_{e\gamma}$ as well as to the other operators. 
For instance scalar and vector operators $Q_{e\vp}$, $O^{(1,3)}_{\vp l}$, and $Q_{\vp e}$ are generated with
\bea
\frac{C_{e\vp}}{\Lambda_{NP}^2} &=&  \frac{1}{2\tilde m^2} X \yr \yr^\dagger X^\dagger y^{SM}_\ell + y^{SM}_\ell \frac{1}{2 m^2} \Xt \yr^\dagger \yr \Xt^\dagger~~~,\\
\frac{C^{(1)}_{\vp l}}{\Lambda_{NP}^2}  &=& - \frac{1}{2 \tilde m^2} X \yr \yr^\dagger X^\dagger~~,~~~~~~~
 C^{(3)}_{\vp l} = 0~~~,~~~~~~~
\frac{C_{\vp e}}{\Lambda_{NP}^2} = \frac{1}{2 m^2} \Xt \yr^\dagger \yr \Xt^\dagger~~~.
\eea
As a result the best probes of the scenario $\yl=0$ are $\mu\to 3e$ , $\mu \to e$ conversion in nuclei and the electron EDM. In the portion of the parameter space where perturbation theory is applicable, LFV allows a compositeness scale close to the TeV,
even in the case of anarchical Yukawas \cite{Feruglio:2015gka}.

Another possibility to reconcile partial compositeness with the TeV scale is to assume an SU(3) flavour symmetry in the strong sector, forcing $m$, $\tilde m$, $\yr$ and $\yl$ to be aligned to the identity matrix~ 
$\mbox{l\hspace{-0.55em}1}$, at least in a suitable basis. If in such a basis we have also $X\propto \mbox{l\hspace{-0.55em}1}$ 
($\tilde X\propto \mbox{l\hspace{-0.55em}1}$), then the unique source of LFV is $\tilde X$ ($X$), which is proportional to the matrix of charged lepton Yukawa couplings. This scenario reproduces the case
of MFV and CLFV are absent in the limit of massless neutrinos \cite{Redi:2013pga}. Even in this MFV limit, sizeable deviations are expected in the Higgs Yukawa couplings.
Indeed the operator  $Q_{e\vp}$ contributes to lepton masses and to higgs couplings with a different weight:
\be
{\cal L}_Y=-{\cal M}_{ij}
~\bar{e}_{Li}\tilde e_{Rj}-
{\cal Y}_{ij}
~h~\bar{e}_{Li}\tilde e_{Rj}+
h.c.~,
\ee
where \be
{\cal M}_{ij}=\left[
y_{\ell ij}^{SM}-\frac{v^2}{2\Lambda_{NP}^2}(C_{e\varphi})_{ij}
\right]
~\frac{v}{\sqrt{2}}~~~,~~~~~~~~{\cal Y}_{ij}=\frac{1}{\sqrt{2}}\left[
y_{\ell ij}^{SM}-\frac{3v^2}{2\Lambda_{NP}^2}(C_{e\varphi})_{ij}
\right]~~~.
\ee
In this case we have $C_{e\varphi}/\Lambda_{NP}^2\approx y^{SM}_\ell \yr\yl/(m \tilde m)$ and
\be
\frac{\delta y_\ell}{y^{SM}_\ell}\approx\frac{v^2}{m\tilde m} \yr \yl~~~,
\ee
at the level of $10\%$ for $m,\tilde m=1$ TeV and $\yl,\yr\approx 1\div 2$, not far from the present sensitivity of LHC to the tau Yukawa coupling.

The future developments of this field are very promising with prospects of exploring new significant territories in the intensity frontier.
Evidence for CLFV and LCPV would  represent a major advance, not only in connection with the flavour structure of NP, but also as a new
step in our quest for a solution of the mystery of fermion masses and mixing angles.

\section*{Acknowledgments}
I would like to thank the organizers of the Rencontres de Blois 2015 for inviting me to this very stimulating workshop and for their very kind hospitality.
I also thank Paride Paradisi and Andrea Pattori for the nice collaboration on which the original part of this talk is based.
This work was supported in part by the Istituto Nazionale di Fisica Nucleare (INFN), by the MIUR-PRIN project 2010YJ2NYW and by the European Union network FP7 ITN INVISIBLES 
(Marie Curie Actions, PITN-GA-2011-289442).

\end{document}